\documentclass[12pt]{iopart}

\usepackage{iopams}
\usepackage{graphicx}

\newcommand{\Fi}[1]{Fig.~\ref{#1}}

\newcommand{\agev}{\mbox{~$A$GeV}}

\newcommand{\rb}[1]{\mbox{\textrm{\scriptsize #1}}}

\newcommand{\sqrts}{\ensuremath{\sqrt{s_{_{\rb{NN}}}}}}
\newcommand{\ybeam}{\ensuremath{y_{\rb{beam}}}}
\newcommand{\lam}{\ensuremath{\Lambda}}
\newcommand{\lab}{\ensuremath{\bar{\Lambda}}}
\newcommand{\myphi}{\ensuremath{\phi}}
\newcommand{\pimin}{\ensuremath{\pi^-}}

\newcommand{\pipm}{\ensuremath{\pi^{\pm}}}
\newcommand{\kmin}{\ensuremath{\textrm{K}^-}}
\newcommand{\kplus}{\ensuremath{\textrm{K}^+}}
\newcommand{\kpm}{\ensuremath{\textrm{K}^{\pm}}}

\newcommand{\kzero}{\ensuremath{\textrm{K}^{0}}}
\newcommand{\kzerob}{\ensuremath{\bar{\textrm{K}}^{0}}}
\newcommand{\sigzero}{\ensuremath{\Sigma^{0}}}
\newcommand{\sigzerob}{\ensuremath{\bar{\Sigma}^{0}}}
\newcommand{\sigpm}{\ensuremath{\Sigma^{\pm}}}
\newcommand{\sigpmb}{\ensuremath{\bar{\Sigma}^{\pm}}}
\newcommand{\xis}{\ensuremath{\Xi^{0,-}}}
\newcommand{\xisb}{\ensuremath{\bar{\Xi}^{0,+}}}
\newcommand{\ommin}{\ensuremath{\Omega^-}}

\newcommand{\omplus}{\ensuremath{\bar{\Omega}^+}}

\newcommand{\mypt}{\ensuremath{p_{\rb{t}}}}

\newcommand{\mt}{\ensuremath{m_{\rb{t}}}}

\newcommand{\meanmtm}{\ensuremath{\langle m_{\rb{t}} \rangle - m_{\rb{0}}}}

\newcommand{\mub}{\ensuremath{\mu_{\rb{B}}}}
\newcommand{\gams}{\ensuremath{\gamma_{\rb{S}}}}

\begin{document}

\title[]{Review of Results from the NA49 Collaboration}

\author{Christoph Blume for the NA49 Collaboration$^a$}

\address{Institut f\"ur Kernphysik der J.W.~Goethe Universit\"at, 
60486 Frankfurt am Main, Germany}

\begin{abstract}
New results of the NA49 collaboration on strange particle production
are presented. Rapidity and transverse mass spectra as well as
total multiplicities are discussed.
The study of their evolution from AGS over SPS to the highest RHIC 
energy reveals a couple of interesting features. These include a 
sudden change in the energy dependence of the \mt-spectra and of the
yields of strange hadrons around 30\agev. Both are found to be difficult
to be reproduced in a hadronic scenario, but might be an indication
for a phase transition to a quark gluon plasma.
\end{abstract}




\section{Introduction}

In the recent years the NA49 experiment has collected data on Pb+Pb 
collisions at beam energies between 20 to 158\agev with the objective 
to cover the critical region of energy densities where the expected 
phase transition to a deconfined phase might occur in the early stage 
of the reactions. 
In this contribution the energy dependence of \mt- and
rapidity-distributions, as well as the production rates of strange
particles are reviewed.
NA49 is a fixed target experiment at the CERN SPS and consists of
a large acceptance magnetic spectrometer equipped with four TPCs as 
tracking devices and a forward calorimeter for centrality selection.
Details on the experimental setup can be found in \cite{na49nim}.

\section{Rapidity spectra}

\begin{figure}[htb]
\begin{center}
\includegraphics[width=110mm]{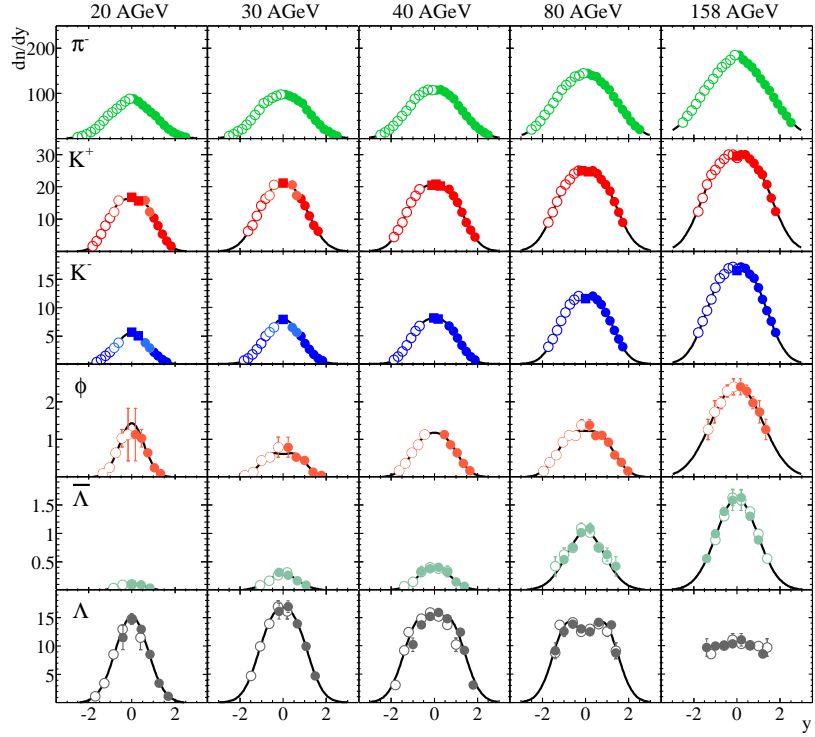}
\end{center}
\caption[]{The rapidity spectra of hadrons produced in central 
Pb+Pb collisions
(7\% at 20-80\agev, 5\% (\pimin, \kplus, \kmin) and
10\% (\myphi, \lab, \lam) at 158\agev). 
The closed symbols indicate measured points, open 
points are reflected with respect to mid-rapidity.
The solid lines represent fits with a single Gaussian or
the sum of two Gaussians.}
\label{rapall}
\end{figure}

The large acceptance of the NA49 spectrometer allows to measure 
particle spectra over a wide range of the longitudinal phase space. 
\Fi{rapall} shows a compilation of the rapidity distributions 
of \pimin, \kplus, \kmin, \myphi, \lab, and \lam. 
The shapes of these distributions are generally well described by
a Gaussian. Only the $\Lambda$-distributions exhibit a strong 
variation of the shape. While at 30\agev\ they are still Gaussian-shaped, 
a clear plateau develops with increasing beam energy.

\begin{figure}[htb]
\begin{center}
\includegraphics[height=65mm]{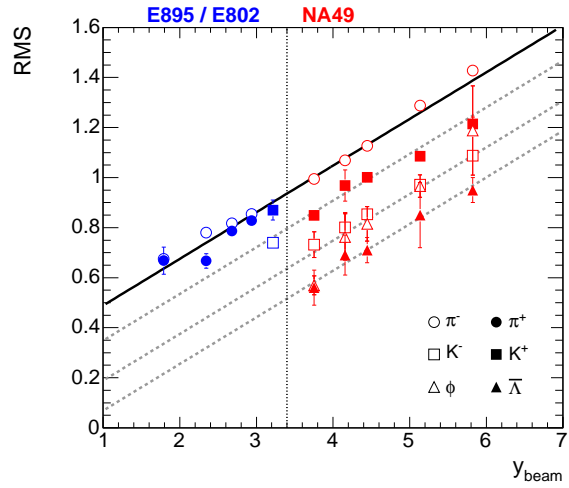}
\end{center}
\caption[]{The RMS values of the rapidity distributions of \pipm, 
\kpm, \myphi, and \lab\ in central Pb+Pb (Au+Au) collisions as a 
function of \ybeam. AGS data are taken from \cite{agspi,agspi2}.
The solid line is a linear fit to the pion data. The dashed
lines have the same slope, but shifted to match the other 
particle species.}
\label{rmsybeam}
\end{figure}

An increase of the RMS-widths of the rapidity spectra, calculated from
the fits shown in \Fi{rapall}, with beam energy 
can be observed which for the pions exhibits to a good approximation a linear 
dependence on the beam rapidity in the center-of-mass system \ybeam\ 
over the whole energy range
covered by the AGS and SPS (see \Fi{rmsybeam}). Between 20\agev\
and 158\agev\ this is also true for the other particle types having a
Gaussian-like distribution, with a clear hierarchy in the widths:
$\sigma(\pimin) > \sigma(\kplus) > \sigma(\kmin) \approx \sigma(\myphi) > \sigma(\lab)$.
However, this seems to break down at lower energies, where the
widths of the kaons apparently approach the ones of the pions.

\begin{figure}[htb]
\begin{center}
\includegraphics[width=1.0\textwidth]{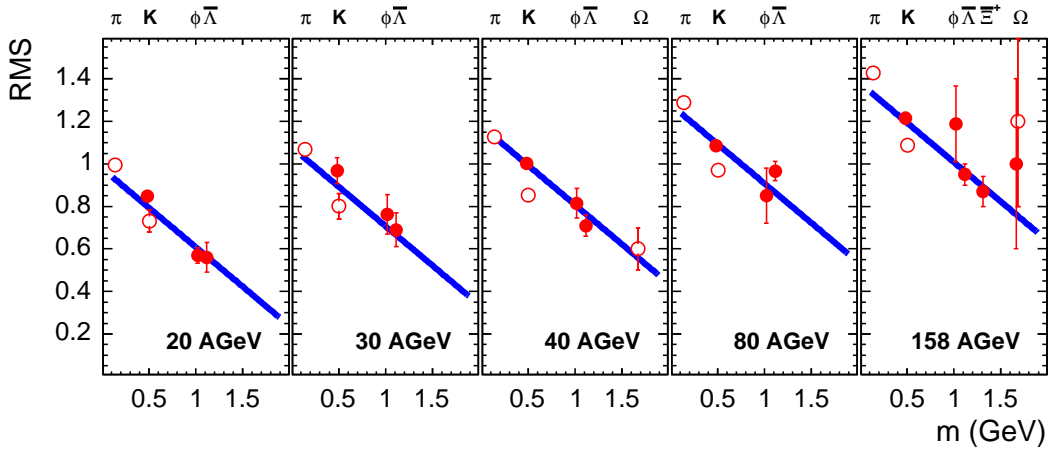}
\end{center}
\caption[]{The RMS-widths of the rapidity distribution as a
function of the particle mass for central Pb+Pb collisions at
different beam energies.
Open symbols indicate negatively charged particles.
The line has always the same slope and are plotted to guide the eye.}
\label{rmsmass}
\end{figure}

The dependence of the RMS-widths on the particle mass at SPS energies
is shown in \Fi{rmsmass}. An approximately linear mass dependence is 
observed, which appears to have the same slope for all SPS-energies. 
This mass dependence can be interpreted as a result of the thermal
spectrum of hadrons superimposed on the longitudinal collective
expansion.

\section{Transverse mass spectra}

\begin{figure}[htb]
\begin{center}
\includegraphics[width=120mm]{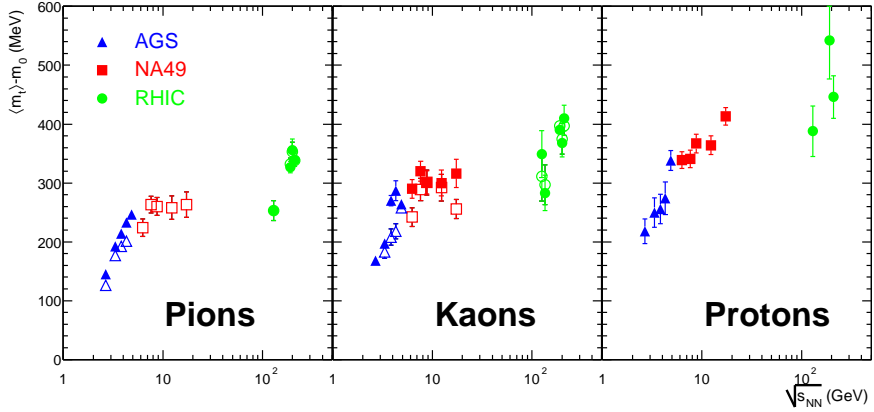}
\end{center}
\caption[]{The energy dependence of \meanmtm\ 
for pions, kaons, and protons at mid-rapidity for 5-10\% most central
Pb+Pb/Au+Au reactions. Open symbols represent negatively charged particles.
}
\label{meanmt}
\end{figure}

The increase with energy of the inverse slope parameter $T$ of the kaon
\mt-spectra, as derived from an exponential fit, exhibits a sharp
change to a plateau around 30\agev\ \cite{marekqm}. Since the kaon \mt-spectra 
-- in contrast to the ones of the lighter pions or the heavier protons 
-- have to a good approximation an exponential shape, the inverse slope 
parameter provides in this case a good characterization of the spectra.
For other particle species, however, the local slope of the spectra depends
on \mt. Instead, the first moment of the \mt-spectra can be used
to study their energy dependence.
The dependence of \meanmtm\ on the center of mass energy \sqrts\ 
is summarized in \Fi{meanmt}. 
The change of the energy dependence around around a beam energy of 
20 -- 30\agev. is clearly visible for pions and kaons.
While \meanmtm\ rises steeply in 
the AGS energy range, the rise is much weaker from the low SPS energies 
on. To a lesser extend this change is also seen for protons.

\section{Particle yields}

\begin{figure}[t]
\begin{center}
\begin{minipage}[b]{70mm}
\begin{center}
\includegraphics[height=120mm]{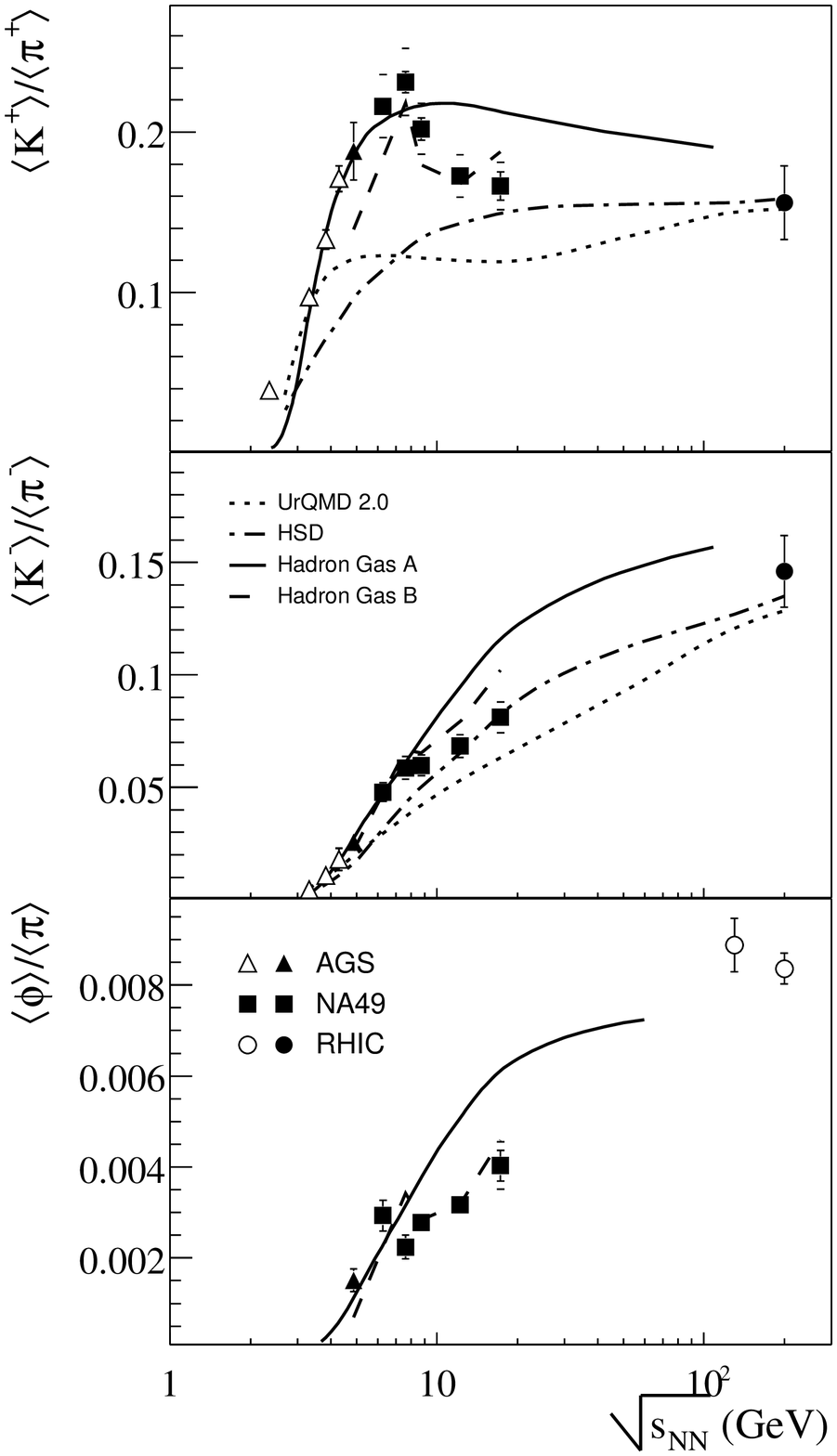}
\end{center}
\end{minipage}
\begin{minipage}[b]{70mm}
\begin{center}
\includegraphics[height=120mm]{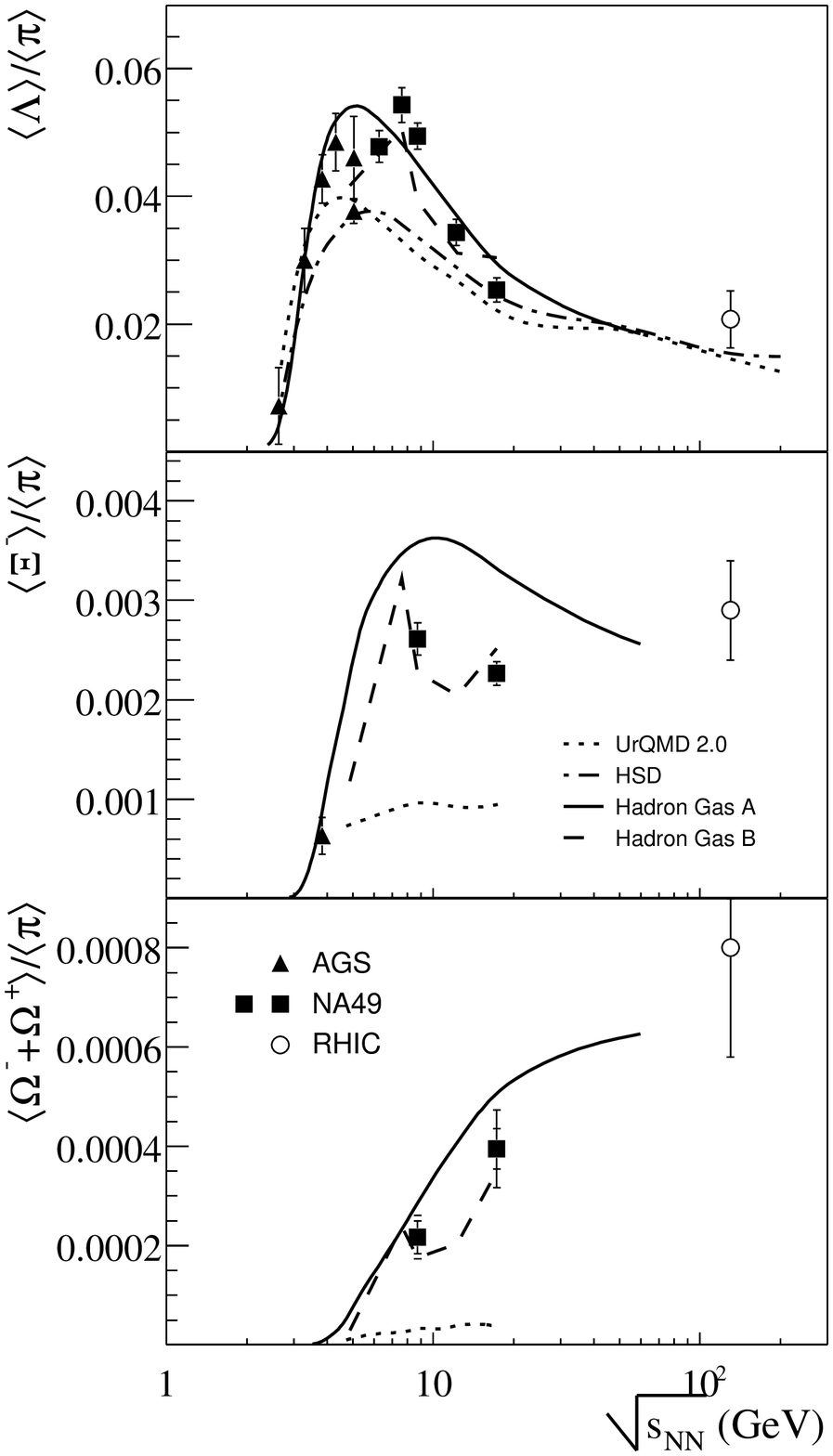}
\end{center}
\end{minipage}
\end{center}
\caption{The energy dependence of the 4$\pi$-yields of
strange hadrons, normalized to the pion yields, in central
Pb+Pb/Au+Au collisions. The data are
compared to string hadronic models \cite{urqmd,hsd} 
(UrQMD 2.0: dotted lines, HSD: dashed-dotted lines) 
and statistical hadron gas models \cite{pbm,becatt}
(with strangeness under-saturation: dashed line, 
assuming full equilibrium: solid line).}
\label{ratios}
\end{figure}

By integrating a measured rapidity distribution, as shown in \Fi{rapall}, 
the total multiplicity of a given particle type can be determined. 
In \Fi{ratios} the energy dependence of the total multiplicities 
for a variety of strange hadrons, normalized to the total
pion yield, is summarized and compared to model predictions. 
A detailed discussion can be found in~\cite{volker}.
Generally, it can be stated that string hadronic models UrQMD and HSD
\cite{urqmd,hsd} do not provide a good description of the data points.
Especially the $\Xi$ and $\Omega$ production is substantially underestimated
and the maximum in the \kmin/\pimin ratio is not reproduced. 
The statistical hadron gas models \cite{pbm,becatt}, on the other hand, 
provide a better overall description of the measurements. However, the 
introduction of an energy dependent strangeness under-saturation factor 
\gams\ is needed \cite{becatt}, in order to capture the structures in the 
energy dependence of most particle species (\kplus, \kmin, \myphi, $\Xi$).

\begin{figure}[htb]
\begin{center}
\begin{minipage}[b]{70mm}
\begin{center}
\includegraphics[height=60mm]{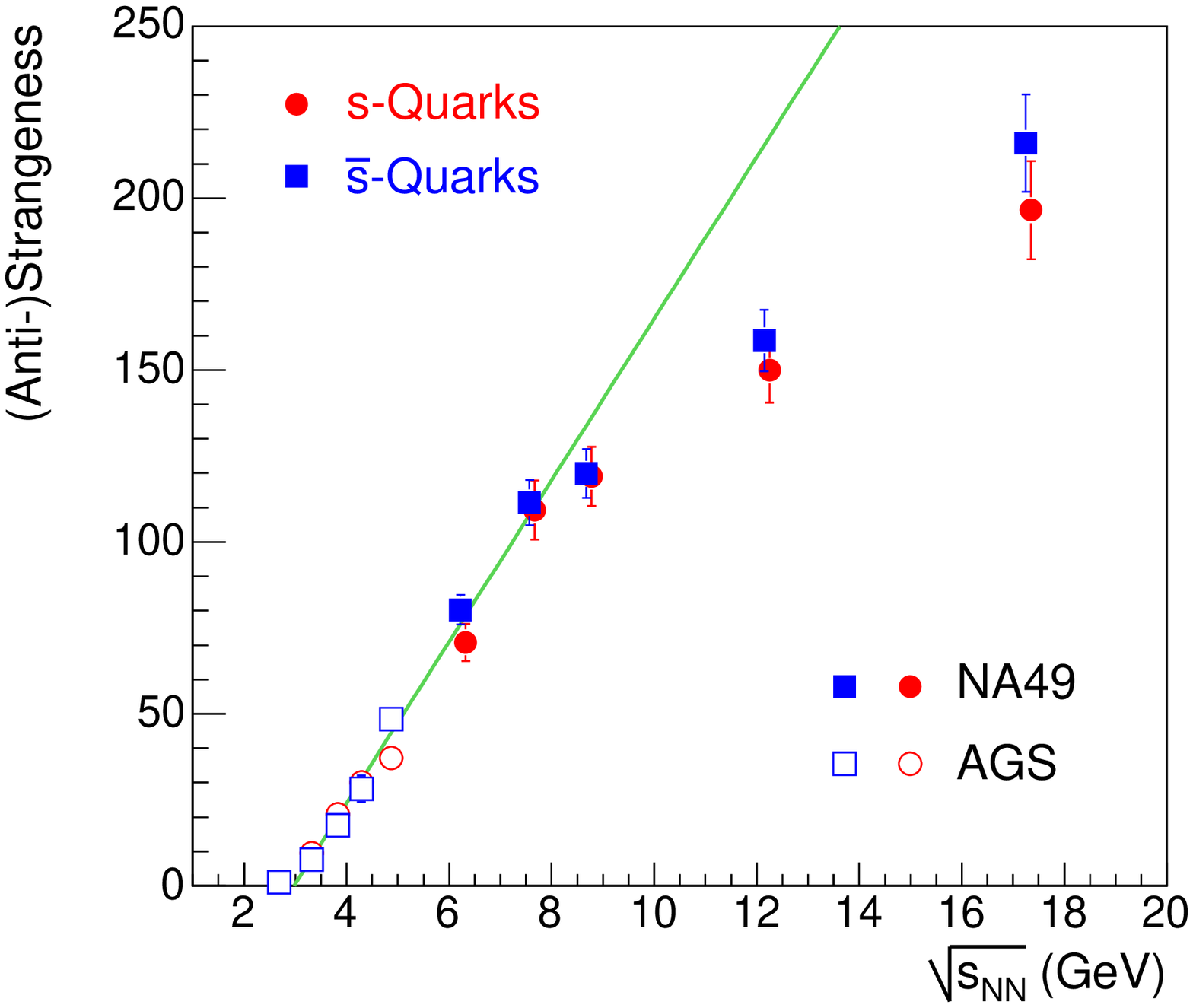}
\end{center}
\end{minipage}
\begin{minipage}[b]{70mm}
\begin{center}
\includegraphics[height=60mm]{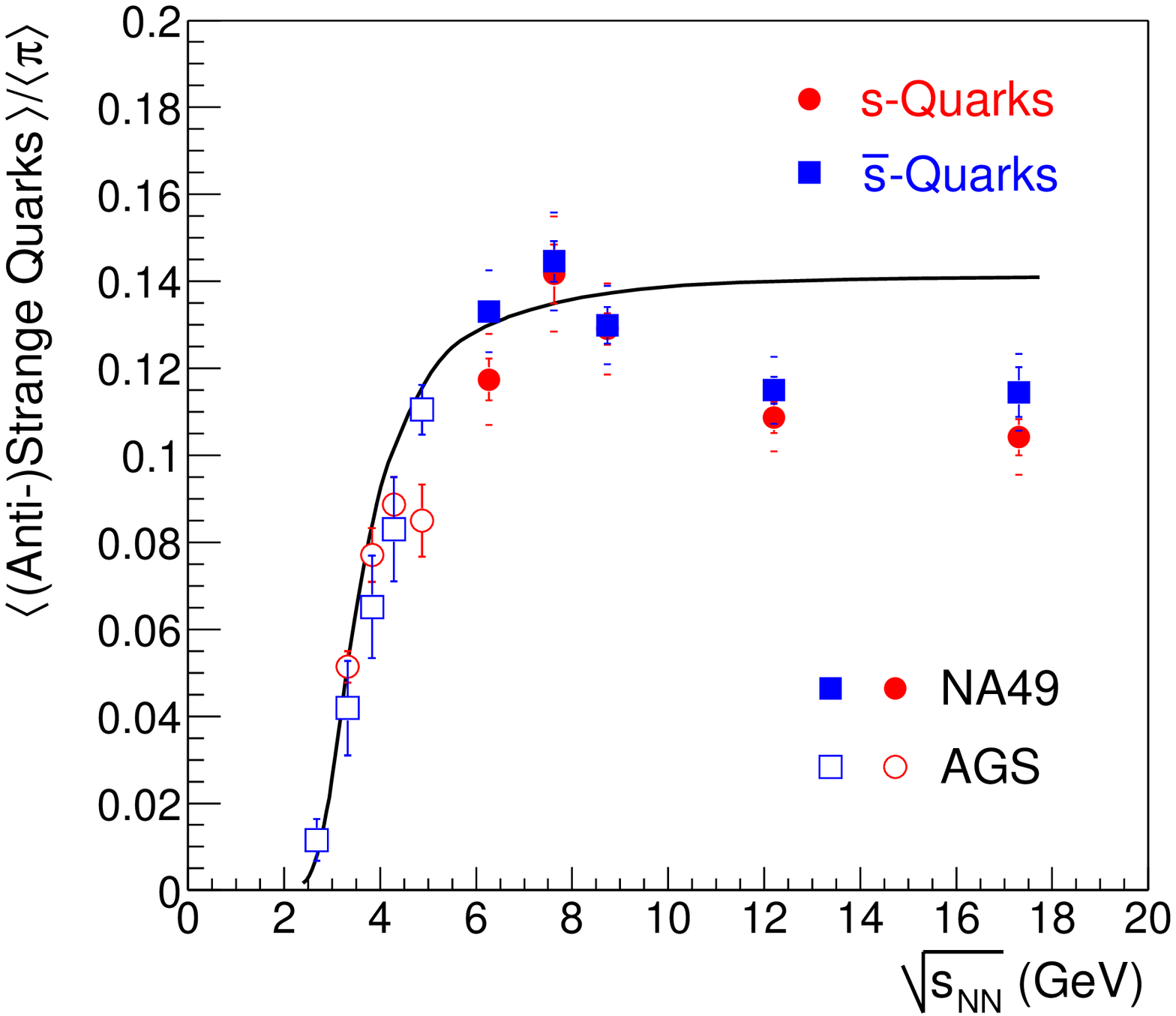}
\end{center}
\end{minipage}
\end{center}
\caption[]{Left: The total number of strange quarks and anti-quarks
as carried by kaons and hyperons versus the collision energy
for central Pb+Pb (Au+Au) reactions. 
The line represents a linear fit to the low energy data.
Right: The ratio of the number of strange aquarks and anti-quarks
to the pion yield as a function of \sqrts.
The solid line represents the prediction of the statistical
hadron gas model with full equilibration of strangeness \cite{redlich}.}
\label{ssbar}
\end{figure}

From the measured total yields of the strange particles the
energy dependence of the number of produced strange quarks
and anti-quarks is constructed (see left panel of \Fi{ssbar}). The strange 
quark carriers which are taken into account are \kmin, \kzero, \lam\ 
(including \sigzero), \xis, \ommin, and \sigpm. For the strange anti-quark 
these are \kplus, \kzerob, \lab\ (including \sigzerob), \xisb, \omplus, and
\sigpmb \footnote{The \kzero\ contribution is calculated using isospin
symmetry ($\langle \kplus \rangle \approx \langle \kzero \rangle$,
$\langle \kmin \rangle \approx \langle \kzerob \rangle$). If no 
measurement is available, the values for the $\Xi$ and $\Omega$ 
yields were taken from statistical model fits \cite{becatt}. 
The \sigpm\ contribution is estimated based on the empirical factor
$(\langle \sigpm \rangle + \langle \lam \rangle)/\langle \lam \rangle = 1.6$
\cite{wrobl}.
Note that the strange quarks from the \myphi\ and $\eta$ are not included.}
The $s$- and $\bar{s}$-yields derived from the
NA49 measurements agree at all energies, illustrating
the consistency of the analysis. A departure from the energy dependence
of strangeness production as observed at lower energies (indicated by the
straight line) is observed around 30\agev. 
If divided by the total number of pions a clear maximum of the relative
strangeness production at the same energy can be seen (right panel of
\Fi{ssbar}). While the statistical hadron gas model with full strangeness
equilibration \cite{redlich} matches the data below this maximum quite well, 
it over-predicts the measurements at higher energies.

\section{Conclusions}

\begin{figure}[htb]
\begin{center}
\includegraphics[height=75mm]{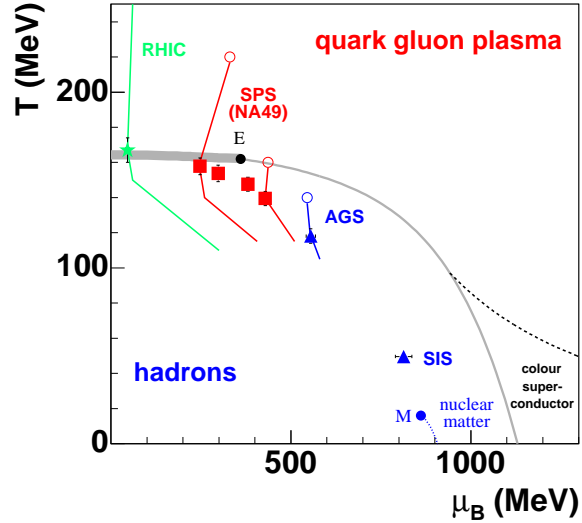}
\end{center}
\caption[]{The phase diagram of strongly interacting matter. The curved lines
represent the phase boundary between a hadron gas and a quark gluon plasma
as expected from lattice calculations. The critical point $E$ is the
endpoint of the first order transition line (thin line on the right side of $E$)
\cite{stephan}.
On the left side of $E$ a smooth cross over is expected. The points are the
chemical freeze-out points as derived from a fit with a statistical hadron
gas model \cite{becatt}. The open symbols schematically indicate possible
initial parameters of the reaction systems, which then might evolve along 
paths as depicted by the vertical lines.}
\label{phasediag}
\end{figure}

The results of the NA49 energy scan program have revealed a variety
of interesting features. Among these are a clear change in the energy
dependence of the \mt-spectra and a maximum in the strangeness to
pion ratio around 30\agev. Both are difficult to explain in a hadronic
scenario, but can be understood as a reflection of a phase transition 
\cite{smes,kodama}.
One of the most remarkable features of the phase diagram, shown
in \Fi{phasediag}, is the fact that the line of the first order
phase transition ends in a second order critical point $E$ 
(for a review of the current theoretical situation see \cite{stephan}).
Its position is subject to large theoretical uncertainties, but
recent lattice calculations \cite{fodor,bielefd} with physical quark masses
indicate that it might be around $\mub = 360$~MeV. If this
estimate holds, it might be possible to access the critical
endpoint with heavy ion reactions at SPS energies. Since the above described 
results on the energy dependence of hadronic observables might indicate that
the phase transition is already reached at beam energies of 30\agev,
where the chemical freeze-out happens at $\mub > 360$~MeV.
This would imply that at this energy the first order transition line
is crossed (see vertical lines in \Fi{phasediag}), 
while at the top SPS energy the evolution passes through the rapid
crossover region. Therefore, a careful scan of this region of beam energy, 
accompanied by an additional variation of the system size, might allow to 
experimentally identify the position of the endpoint.
An indication for the endpoint would possibly be a strong increase of
fluctuations in e.g. transverse momentum, particle ratios or multiplicities.
So far, only limited attempts of systematically studying these effects 
in the interesting energy range have been made. 
Significant non-statistical fluctuations of the kaon to pion ratio
at low SPS energies are reported in \cite{roland}. Also, large multiplicity
and \mypt fluctutions in collisions of small systems at 158\agev have
been observed \cite{marekqm}. However, these effects are still far from
having a clear connection to the critical point. Future measurements
that would scan the interesting energy range and the system size more 
closely, might have a chance of seeing stronger evidence for its existence.

\section*{Acknowledgments}

This work was supported by the US Department of Energy
Grant DE-FG03- \\
97ER41020/A000,
the Bundesministerium fur Bildung und Forschung, Germany, 
the Polish State Committee for Scientific Research 
(2 P03B 130 23, SPB/CERN/P-03/Dz 446/2002-2004, 2 P03B 04123), 
the Hungarian Scientific Research Foundation (T032648, T032293, T043514),
the Hungarian National Science Foundation, OTKA, (F034707),
the Polish-German Foundation, and the Korea Research Foundation 
Grant (KRF-2003-070-C00015).

\section*{Notes}
\begin{itemize}
\item[a] The NA49 collaboration:\\
{\small
C.~Alt$^{9}$, T.~Anticic$^{21}$, B.~Baatar$^{8}$,D.~Barna$^{4}$,
J.~Bartke$^{6}$, 
L.~Betev$^{9,10}$, H.~Bia{\l}\-kowska$^{19}$, A.~Billmeier$^{9}$,
C.~Blume$^{9}$,  B.~Boimska$^{19}$
% ...
}\\
{\footnotesize
% ...
$^{4}$KFKI Research Institute for Particle and Nuclear Physics, Budapest, Hungary.
% ...
$^{6}$Institute of Nuclear Physics, Cracow, Poland.
% ...
$^{8}$Joint Institute for Nuclear Research, Dubna, Russia.
$^{9}$Fachbereich Physik der Universit\"{a}t, Frankfurt, Germany.
$^{10}$CERN, Geneva, Switzerland.
% ...
$^{19}$Institute for Nuclear Studies, Warsaw, Poland.
% ...
$^{21}$Rudjer Boskovic Institute, Zagreb, Croatia.
}
\end{itemize}



\end{document}